\documentclass[pre,floatfix,twocolumn,showpacs]{revtex4}
\usepackage{amsmath}
\usepackage{amsfonts}
\usepackage{mathrsfs}
\usepackage{epsfig}
\usepackage{graphicx}
\usepackage{dcolumn}
\usepackage{amsmath}

\def\be{\begin{equation}}
\def\ee{\end{equation}}

\begin{document}

\title{Epidemiological Dynamics of the 2009 Influenza A(H1N1)v
Outbreak in India}

\author{T Jesan$^{1,2}$, Gautam I Menon$^1$ and Sitabhra
Sinha$^{1,*}$}
\affiliation{
$^1$ The Institute of Mathematical Sciences, CIT Campus, Taramani,
Chennai 600113, India\\
$^2$ Health Physics Division, Bhaba Atomic Research Center, Kalpakkam
603102, India\\
$^*$ Corresponding author, E-mail: sitabhra@imsc.res.in	
}
\date{\today}
\begin{abstract}
We analyze the time-series data for the onset of A(H1N1)v influenza
pandemic in India during the period June 1- September 30, 2009. Using
a variety of statistical fitting procedures, we obtain a robust
estimate of the exponential growth rate $\langle \lambda \rangle \simeq
0.15$. This corresponds to a
basic reproductive number $R_0 \simeq 1.45$ for influenza A(H1N1)v in India, a
value which lies towards the lower end of the range of values reported
for different countries affected by the pandemic.

\noindent
\end{abstract}
\pacs{05.45.Tp,87.19.xd,02.50.-r,87.23.Cc}
\maketitle
\newpage

\section{Introduction}
A novel influenza strain termed influenza A(H1N1)v, first identified
in Mexico in March 2009, has rapidly spread to different countries and
is currently the predominant influenza virus in circulation
worldwide~\cite{flu09,Jameel10}.
As of April 11, 2010, it has caused at least 17798 deaths in 214
countries~\cite{WHO10}. The first confirmed case in India, a passenger arriving
from the USA, was detected on May 16, 2009 in Hyderabad. The initial
cases were passengers arriving by international flights. However,
towards the end of July, the infections appeared to have spread into
the resident population with an increasing number of cases being
reported for people who had not been abroad. As of 11 April 2010,
there have been 30352 laboratory confirmed cases in India (out of
132796 tested) and 1472 deaths have been reported, i.e., $5~\%$ of the
cases which tested positive for influenza A(H1N1)v~\cite{MHO10}. 

To devise effective strategies for combating the spread of pandemic
influenza
A(H1N1), it is essential to estimate the transmissibility of this
disease in a reliable manner. This is generally characterized by the
reproductive number $R$, defined as the average number of secondary
infections resulting from a single (primary) infection. A special case
is the basic reproductive number $R_0$, which is the value of $R$ measured
when the overall population is susceptible to the infection as is the
case at the initial stage of an epidemic. Estimate of the basic
reproduction number for influenza A(H1N1)v in reports published from
data obtained for different countries vary widely. For example, $R_0$ has
been variously estimated to be between 2.2 to 3.0 for
Mexico~\cite{Boelle09}, 1.72
for Mexico City~\cite{Cruz-Pacheco09}, between 1.4 and 1.6 for La
Gloria in Mexico~\cite{Fraser09},
between 1.3 to 1.7 for the United States~\cite{Yang09} and 2.4 for Victoria
State in Australia~\cite{McBryde09}. The divergence in the estimates for the basic
reproductive number may be a result of under-reporting in the early
stages of the epidemic or due to climatic variations. They may also
possibly reflect the effect of different control strategies used in
different regions, ranging from social distancing such as school
closures and confinement to antiviral treatments.

In this paper, we estimate the basic reproductive number for the
infections using the time-series of infections in India extracted from
reported data. By assuming an exponential rise in the number of
infected cases $I(t)$ during the initial stage of the epidemic when most
of the population is susceptible, we can express the basic
reproductive number as $R_0 = 1+\lambda \tau$ (see, e.g.,
Ref.~\cite{Anderson92},
p. 19), where $\lambda$
is the rate of exponential growth in the number of infections, and
$\tau$
is the mean generation interval, which is approximately equal to 3
days~\cite{Cruz-Pacheco09}. Using the time-series data we obtain the slope 
$\lambda$ of the
exponential growth using several different statistical techniques. Our
results show that this quantity has a value of around 0.15,
corresponding to $R_0 \simeq 1.45$.

\section{Methods}
We used data from the daily situation updates available from the
website of the Ministry of Health and Family Welfare, Government of
India~\cite{mohfw}. In our analysis, data up to September 30, 2009 was used,
corresponding to a total of 10078 positive cases. Note that, after
September 30, 2009, patients exhibiting mild flu like symptoms
(classified as categories A and B) were no longer tested for the
presence of the influenza A(H1N1) virus.

As the data exhibit very large fluctuations, with some days not
showing a single case while the following days show extremely large
number of cases, it is necessary to smooth the data using a moving
window average. We have used an $n$-day moving average ($n=2-10$), which
removes large fluctuations while remaining faithful to the overall
trend. 

\begin{figure*}
\centering
\includegraphics[width=0.99\linewidth,clip]{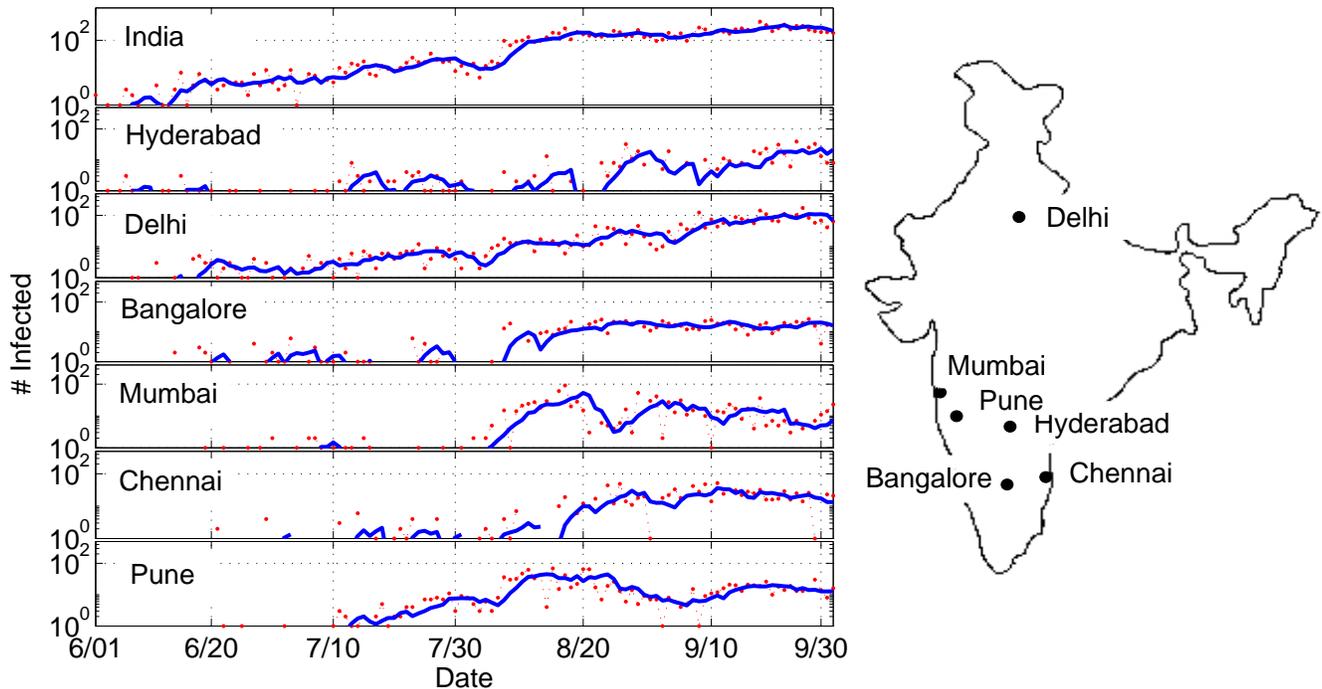}
\caption{
Time-series of the number of infected cases, $\#$Infected, of
influenza A(H1N1)v showing the daily data (dotted) as well as the
5-day moving average (solid line) for India and the six metropolitan
areas with the highest number of infections (whose geographic
locations are shown in the adjoining map). The period shown is from
June 1 to September 30, 2010. At the beginning of this period most of
the infected people were arriving from abroad, while at the end of it
the infection was entrenched in the local population. The data shows
that almost all the cities showed a simultaneous increase in the
number of infections towards the end of July and the beginning of
August. This is manifested as a sudden rise in $\#$Infected for India as
a whole (note the semilogarithmic scale), and can be taken as the
period in which the infection started spreading in the resident
population.}
\label{Figure1}       
\end{figure*}
\section{Results}
The incidence data for the 2009 pandemic influenza data in India
immediately reveals that the disease has been largely confined to the
urban areas of the country. Indeed, 6 of the 7 largest metropolitan
areas of India (which together accommodate about 5~\% of the Indian
population~\cite{worldgazetteer}) account for 7139 infected cases up to September 30,
2009, i.e., $70.8~\%$ of the data-set we have used.

Figure~\ref{Figure1} shows the daily number of confirmed infected cases, as well
as,
the 5-day moving average from June 1 to September 30, 2009,
for the country as a whole and the six major metropolitan
areas which showed the highest incidence of the disease: Hyderabad,
Delhi, Bangalore, Mumbai, Chennai and Pune. The adjoining map shows
the geographic locations of these six cities. In the period up to July
2009, infections were largely reported in people arriving from abroad.
There is a marked increase in the number of infections towards the end
of July and the beginning of August 2009 in all of these cities (note
that the ordinate is in logarithmic scale). This is manifest as a
sudden rise in the number of infected cases for the country as a
whole, implying that the infection started spreading in the resident
population in the approximate period of 28 July to August 12. 

\begin{figure}
\centering
\includegraphics[width=0.99\linewidth,clip]{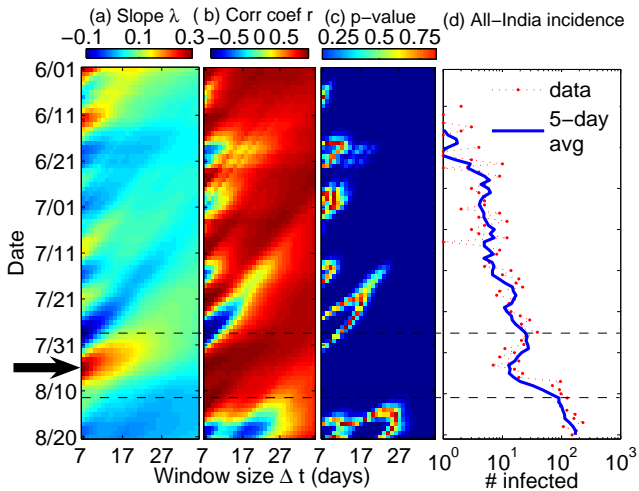}
\caption{
(a) The exponential slope $\lambda$ estimated from the time-series
data of number of infected cases, $\#$Infected, averaged over a 5-day
period to smoothen the fluctuations (d, solid curve). The slope $\lambda$
is calculated by considering the number of infected cases over a
moving window having different sizes ($\Delta t$), ranging between 7 days
and 36 days. By moving the starting point of the window across the
period 1st June-20th August (in steps of 1 day) and calculating the
best fit linear slope of the data on a semi-logarithmic scale (i.e.,
time in normal axis, number of infections in logarithmic axis) we
obtain an estimate of $\lambda$. The arrow indicates the region between July
28-August 12 (region within the broken lines), which shows the largest
increase in number of infections within the period under study,
corresponding to the period when the epidemic broke out in the
resident population. Over this time-interval, the average of $\lambda$ is
calculated for the set of starting dates and window sizes over which
(b) the correlation coefficient $r$ between log($\#$Infected) and $t$,
is greater than
$r_{cutoff}$ (we consider $0.75 < r_{cutoff} < 1$ in our analysis) and (c), the
measure of significance for the correlation $p < 0.01$.}
\label{Figure2}       
\end{figure}
Figure~\ref{Figure2}~(a) shows the exponential slope $\lambda$ estimated in the
following way. The time-series of the number of infections is first
smoothed by taking a 5-day moving average. The resulting smoothed
time-series is then used to estimate $\lambda$ by a regression procedure
applied to the logarithm of the number of infected cases
[log($\#$infected)] across a moving window of length $\Delta t$ days. 
The origin
of the window is varied across the period 1st June to 20th August (in
steps of 1 day).
We then repeat the procedure by varying the length of the window over
the range of 7 days to 36 days. To quantify the quality of regression we
calculate the correlation coefficient $r$ [Fig.~\ref{Figure2}~(b)] between log
($\#$Infected) and time (in days), and its measure of significance $p$
[Fig.~\ref{Figure2}~(c)]. The correlation coefficient $r$ is bounded between $-1$ and
1, with a value closer to 1 indicating a good fit of the data to an
exponential increase in the number of infections. The measure of
significance of the fitting is expressed by the corresponding $p$-value,
which expresses the probability of obtaining the same correlation by
random chance from uncorrelated data. The average of the estimated
exponential slope $\lambda$ is obtained by taking the mean of all
values of $\lambda$
obtained for windows originating between July 28-Aug 12 and of various
sizes, for which the correlation coefficient $r > r_{cutoff}$ (we consider
$0.75<r_{cutoff} <1$ in our analysis) and the measure of significance
$p < 0.01$. For comparison, we show again in Figure~\ref{Figure2}~(d) the number of
infected cases of H1N1 in India (dotted) together with its 5-day
moving average (solid line). The horizontal broken lines running
across the figure indicate the period
between July 28 and August 12 which exhibited the highest increase in
number of infections within the period under study (from 1st June to
30th September) .

\begin{figure}
\centering
\includegraphics[width=0.99\linewidth,clip]{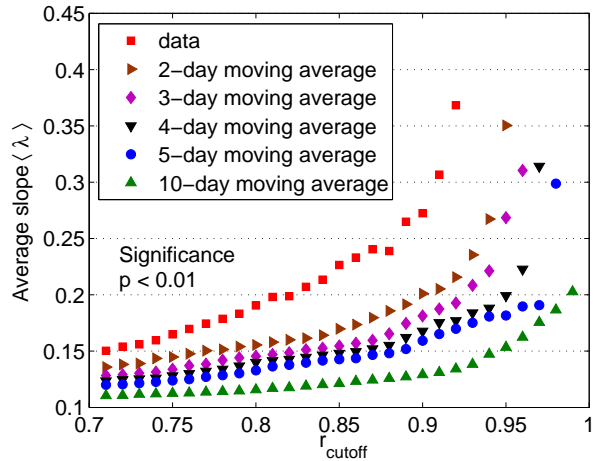}
\caption{
Average slope $\langle \lambda \rangle$ of the variation in log($\#$Infected) 
with
time $t$, as a function of the threshold of correlation coefficient,
$r_{cutoff}$, used to filter the data. The averaging is performed for
infections occurring within the period July 28-August 12 (for details
see caption to Fig.~\ref{Figure2}). Different symbols indicate the actual daily
time-series data (squares) and the data smoothed over a moving $n$-day
period, with $n$ = 2 (right-pointed triangle), 3 (diamond), 4 (inverted
triangle), 5 (circle) and 10 (triangle). The significance of the
correlation between log($\#$Infected) with time $t$, $p<0.01$ for all data
points used in performing the average. Note that for $n=3,4,5$ the data
show very similar profiles for variation of $\langle \lambda \rangle$
with $r_{cutoff}$,
indicating the robustness of the estimate with respect to different
values of $n$ used. The sudden increase in the value of the average
slope around $r_{cutoff} \simeq 0.9$ implies that beyond this region the slope
depends sensitively on the cutoff value. Considering the region where
the variation is more gradual gives us an approximate value of the
slope $\lambda \sim 0.15$, corresponding to a basic reproduction
number $R_0 \simeq 1.45$.}
\label{Figure3}       
\end{figure}
Figure~\ref{Figure3} shows the average exponential slope $\langle \lambda \rangle$
as a function of
$r_{cutoff}$, calculated for the original data and for different
periods $n$
over which the moving average is taken ($n = 2,3,4,5$ and 10). For $n
=$
3-5, the data show a similar profile indicating the robustness of the
estimate of the average exponential slope $\langle \lambda \rangle$
with respect to
different values of $n$. The sudden increase in $\langle \lambda
\rangle$ around $r_{cutoff} \simeq 0.9$ implies that beyond this
region the slope depends sensitively on
the cutoff value. Considering the region where the variation is
smoother gives an approximate value $\lambda \sim 0.15$, corresponding
to a basic
reproductive number for the epidemic $R_0 = 1+\lambda\tau \simeq 1.45$, 
assuming the mean generation interval, $\tau= 3$ days.

We compute the confidence bounds for the estimate of $R_0$ from the 5-day
moving average time-series by using the {\em confint} function of
the scientific software MATLAB~\cite{MATLAB}.
This function generates the goodness of fit statistics using the
solution of the least squares fitting of log($\#$Infected) to a linear
function. It results in a mean
value $\langle \lambda
\rangle = 0.16$, with the corresponding $95~\%$ confidence intervals
calculated as
[0.116, 0.206], consistent with our previous estimate of $R_0 \simeq
1.45$.

\begin{figure}
\centering
\includegraphics[width=0.99\linewidth,clip]{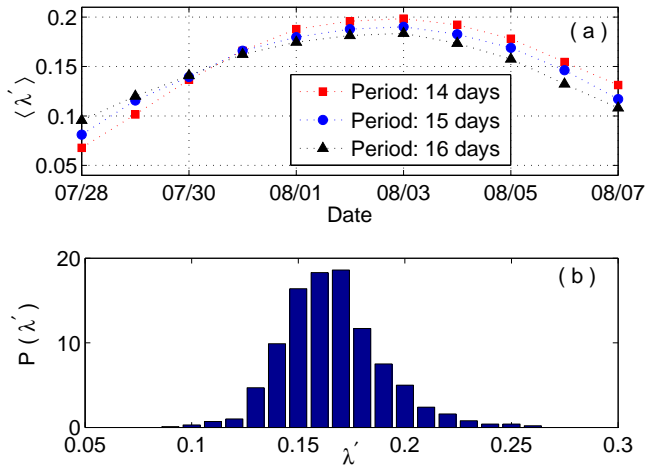}
\caption{
(a) The averages of the bootstrap estimates for the
exponential slope, $\lambda^{\prime}$, calculated for different periods 
(with the
abscissa indicating the starting date and the symbol indicating the
duration) from the 5-day moving average time-series data of infected
cases in India. The curves corresponding to the periods of different
durations (14-16 days) intersect around July 31, 2010, indicating that
the value of the average exponential slope is relatively robust with
respect to the choice of the period about this date. 
(b) The
distribution of bootstrap estimates of the exponential slope for the
period July 31 to August 15, 2009. The average slope $\langle
\lambda^{\prime} \rangle$ obtained
from 1000 bootstrap samples is 0.166 with a standard deviation of
0.024, which agrees with the approximate value of $\lambda= 0.15$
(corresponding to $R_0 = 1.45$) calculated in Fig.~\ref{Figure3}.}
\label{Figure4}       
\end{figure}
We have also used bootstrap methods to estimate the exponential slope,
$\lambda$. This involves selecting random samples with replacement from the
data such that the sample size equals the size of the actual data-set.
The same analysis that was performed on the empirical data is then
repeated on each of these samples. The range of the estimated values
$\lambda^{\prime}$ calculated from the random samples allows determination 
of the
uncertainty in estimation of $\lambda$. Fig.~\ref{Figure4}~(a) shows the average,
$\langle \lambda^{\prime} \rangle$,
calculated for different periods (with abscissa indicating the
starting date and the symbol indicating the duration of the period)
from the 5-day moving average time-series data of infected cases. The
curves corresponding to the periods of different durations (14-16
days) intersect around July 31, 2010, indicating that the value of the
average exponential slope is relatively robust with respect to the
choice of the period about this date. The average value of the
bootstrap estimates $\lambda^{\prime}$ at the intersection of the three 
curves is
0.15, in agreement with our earlier calculations of $\lambda$. 

Fig.~\ref{Figure4}~(b) shows the distribution of the bootstrap estimates of the
exponential slope for a particular period, July 31 to August 15, 2009.
The average slope $\langle \lambda^{\prime} \rangle$ obtained from 1000 bootstrap samples for this
period is 0.166 with a standard deviation of 0.024, which indicates
that the spread of values around the average estimate of $\langle
\lambda^{\prime} \rangle$ = 0.15
is not large. This confirms the reliability of the
estimated value of the exponential slope, and hence of our calculation
of the basic reproductive number.

\section{Discussion}
It may appear surprising that there was a very high number of
infections in Pune (1238 positive cases up to September 30), despite
it being less well-connected to the other major metropolitan cities of
India, in comparison to urban centres that did not show a high
incidence of the disease. For example, the Kolkata metropolitan area,
which has a population around three times the population of the Pune
metropolitan area~\cite{worldgazetteer}, had only 113 positive cases up to September
30. This could possibly reflect the role of local climatic conditions:
Pune, located at a relatively higher altitude, has a generally cooler
climate than most Indian cities. In addition, the close proximity of
Pune to Mumbai and the high volume of road traffic between these two
cities could have helped in the transmission of the disease. Another
feature pointing to the role of local climate is the fact that in
Chennai, most infected cases were visitors from outside the city,
while in Pune, the majority of the cases were from the local
population, even though the total number of infected cases listed for
the two cities in our data-set are comparable (928 in Chennai and 1213
in Pune). This suggests the possibility that the incidence of the
disease in Pune could have been aided by its cool climate, in contrast
to the hotter climate of the coastal city of Chennai.

The calculation of $R_0$ for India assumes well-mixing of the
population (i.e., homogeneity of the contact structure) among the
major cities in India. Given the rapidity of travel between the
different metropolitan areas via air and rail, this may not be an
unreasonable assumption. However, some local variation in the development
of the epidemic in different regions can indeed be seen
(Fig.~\ref{Figure1}) .
Around the end of July, almost all the cities under investigation
showed a marked increase in the number of infected cases - indicating
spread of the epidemic in the local population. This justifies our
assumption of well-mixing in the urban population over the entire
country for calculating the basic reproductive number.

To conclude, we stress the implications of our finding that the basic
reproductive number for pandemic influenza A(H1N1)v in India lies
towards the lower end of the values reported for other affected
countries. This suggests that season-to-season and country-to-country
variations need to be taken into account in order to formulate
strategies for countering the spread of the disease. Evaluation of
the reproductive number, once control measures have been initiated, is
vital in determining the future pattern of spread of the disease.

\vspace{0.5cm}
\noindent
{\bf Acknowledgements}

We acknowledge the IMSc Complex Systems project and PRISM, IMSc for support.


\end{document}